\documentclass{aa}  
\usepackage{graphicx, siunitx}
\usepackage{txfonts}
\usepackage{threeparttable}

\usepackage{hyperref}
\usepackage{float}
\usepackage{xcolor}
\usepackage{lineno}

\def\CIVdbl{{\rm C~}\kern 0.1em{\sc iv}~$\lambda\lambda 1548, 1550$}
\def\MgIIdbl{{\rm Mg~}\kern 0.1em{\sc ii}~$\lambda\lambda 2796, 2803$}
\def\NVdbl{{\rm N}\kern 0.1em{\sc V}~$\lambda\lambda 1238, 1242$}  
\def\OVIdbl{{\rm O}\kern 0.1em{\sc vi}~$\lambda\lambda 1031, 1037$}
\def\SiIVdbl{{\rm Si~}\kern 0.1em{\sc iv}~$\lambda\lambda 1394, 1403$}
\def\AlIIIdbl{{\rm Al~}\kern 0.1em{\sc iii}~$\lambda\lambda 1855, 1863$}
\def\FeIIdbl{{\rm Fe~}\kern 0.1em{\sc ii}~$\lambda\lambda 2383, 2600$}

\def\AlII{\hbox{{\rm Al~}\kern 0.1em{\sc ii}}}
\def\AlI{\hbox{{\rm Al~}\kern 0.1em{\sc i}}}
\def\AlIII{\hbox{{\rm Al~}\kern 0.1em{\sc iii}}}
\def\CaI{\hbox{{\rm Ca}\kern 0.1em{\sc i}}}
\def\CaII{\hbox{{\rm Ca}\kern 0.1em{\sc ii}}}
\def\CrII{\hbox{{\rm Cr}\kern 0.1em{\sc ii}}}

\def\CI{\hbox{{\rm C~}\kern 0.1em{\sc i}}}
\def\CII{\hbox{{\rm C~}\kern 0.1em{\sc ii}}}
\def\CIII{\hbox{{\rm C~}\kern 0.1em{\sc iii}}}
\def\CIV{\hbox{{\rm C~}\kern 0.1em{\sc iv}}}
\def\CV{\hbox{{\rm C}\kern 0.1em{\sc v}}}

\def\HI{\hbox{{\rm H~}\kern 0.1em{\sc i}}}
\def\HII{\hbox{{\rm H~}\kern 0.1em{\sc ii}}}
\def\Lya{\hbox{{\rm Ly}\kern 0.1em$\alpha$}}
\def\Lyb{\hbox{{\rm Ly}\kern 0.1em$\beta$}}
\def\Lyg{\hbox{{\rm Ly}\kern 0.1em$\gamma$}}
\def\Lyfive{\hbox{{\rm Ly}\kern 0.1em$5$}}
\def\Lysix{\hbox{{\rm Ly}\kern 0.1em$6$}}
\def\Lyseven{\hbox{{\rm Ly}\kern 0.1em$7$}}
\def\Lyeight{\hbox{{\rm Ly}\kern 0.1em$8$}}
\def\Lynine{\hbox{{\rm Ly}\kern 0.1em$9$}}
\def\Lyten{\hbox{{\rm Ly}\kern 0.1em$10$}}
\def\HeI{\hbox{{\rm He}\kern 0.1em{\sc i}}}
\def\HeII{\hbox{{\rm He}\kern 0.1em{\sc ii}}}
\def\FeI{\hbox{{\rm Fe~}\kern 0.1em{\sc i}}}
\def\FeII{\hbox{{\rm Fe~}\kern 0.1em{\sc ii}}}
\def\FeIII{\hbox{{\rm Fe~}\kern 0.1em{\sc iii}}}
\def\MnII{\hbox{{\rm Mn}\kern 0.1em{\sc ii}}}
\def\MgI{\hbox{{\rm Mg~}\kern 0.1em{\sc i}}}
\def\MgII{\hbox{{\rm Mg~}\kern 0.1em{\sc ii}}}
\def\MgIII{\hbox{{\rm Mg~}\kern 0.1em{\sc iii}}}
\def\MgIV{\hbox{{\rm Mg~}\kern 0.1em{\sc iv}}}
\def\NaI{\hbox{{\rm Na}\kern 0.1em{\sc i}}}
\def\NV{\hbox{{\rm N}\kern 0.1em{\sc V}}}
\def\NII{\hbox{{\rm N}\kern 0.1em{\sc ii}}}
\def\NIII{\hbox{{\rm N}\kern 0.1em{\sc iii}}}

\def\OVI{\hbox{{\rm O}\kern 0.1em{\sc vi}}}
\def\OIV{\hbox{{\rm O}\kern 0.1em{\sc IV}}}
\def\OI{\hbox{{\rm O}\kern 0.1em{\sc i}}}
\def\OII{\hbox{{\rm O}\kern 0.1em{\sc ii}}}
\def\OIII{\hbox{{\rm O}\kern 0.1em{\sc iii}}}
\def\PV{\hbox{{\rm P}\kern 0.1em{\sc v}}}
\def\SiII{\hbox{{\rm Si~}\kern 0.1em{\sc ii}}}
\def\SiIII{\hbox{{\rm Si~}\kern 0.1em{\sc iii}}}
\def\SiIV{\hbox{{\rm Si~}\kern 0.1em{\sc IV}}}
\def\SII{\hbox{{\rm S}\kern 0.1em{\sc ii}}}
\def\SIII{\hbox{{\rm S}\kern 0.1em{\sc iii}}}
\def\SIV{\hbox{{\rm S}\kern 0.1em{\sc iv}}}
\def\SVI{\hbox{{\rm S}\kern 0.1em{\sc vi}}}
\def\TiII{\hbox{{\rm Ti}\kern 0.1em{\sc ii}}}
\def\ZnII{\hbox{{\rm Zn}\kern 0.1em{\sc ii}}}
\def\kms{\hbox{km~s$^{-1}$}}

\def\NHCfmax{$N_{\mathrm{H}}^{C_f^{max}}$}
\def\NHCfmin{$N_{\mathrm{H}}^{C_f^{min}}$}
\def\EkinCfmin{$\dot{E}_{\mathrm{K,EHVO}}^{C_f^{\min}}$}
\def\EkinCfmax{$\dot{E}_{\mathrm{K,EHVO}}^{C_f^{\max}}$}
\def\Nion{\(N_\mathrm{ion}\)}
\def\NH{\(N_\mathrm{H}\)}

\def\lbol{$L\rm_{Bol}$}

\def\edd{$\lambda\rm_{Edd}$}

\def\lsim{\mathrel{\rlap{\lower4pt\hbox{\hskip1pt$\sim$}}
    \raise1pt\hbox{$<$}}}               
\def\gsim{\mathrel{\rlap{\lower4pt\hbox{\hskip1pt$\sim$}}
    \raise1pt\hbox{$>$}}}

\begin{document}

\title{An Extremely-High Velocity Outflow in SMSS J2157-3602, the most luminous quasar in the first 1.3 Gyr}

   \author{Giustina Vietri \inst{\ref{inst8}}\thanks{E-mail: giustina.vietri@inaf.it}
   \and
   Paola Rodr\'iguez Hidalgo 
          \inst{\ref{inst1}}
       \and   
          Amy Rankine\inst{\ref{inst2}}
         \and
        Luca Zappacosta \inst{\ref{inst9}}
        \and
        Enrico Piconcelli \inst{\ref{inst9}}
          \and
        Liliana Flores 
        \inst{\ref{inst1}}
        \and
Ivano Saccheo \inst{\ref{inst9},\ref{inst19}}
\and
Andrea Melandri  \inst{\ref{inst9}}
\and 
Vincenzo Testa  \inst{\ref{inst9}}
\and  
Patrick B. Hall\inst{\ref{inst13}}  
\and 
Flaminia Sarnari\inst{\ref{inst17}}
\and
Wendy F. Garc\'ia Naranjo\inst{\ref{inst1}}
\and  
Tzitzi Romo P\'erez\inst{\ref{inst1}}  
\and
Valentina D’Odorico \inst{\ref{inst12},\ref{inst18}} 
\and   
Giorgio Lanzuisi\inst{\ref{inst14}}  
\and  
Toru Misawa \inst{\ref{inst16}}
\and  
Christopher A. Onken\inst{\ref{inst10},\ref{inst11}}
\and  
Cristian Vignali  \inst{\ref{inst15}}
\and  
Christian Wolf\inst{\ref{inst10},\ref{inst11}}}
\institute{INAF - Istituto di Astrofisica Spaziale e Fisica cosmica Milano, Via Alfonso Corti 12, 20133, Milano, Italy \label{inst8}
   \and
   Physical Sciences Division,
            School of STEM,
            University of Washington Bothell,
            Bothell WA, 98011, USA \label{inst1}
         \and
             Institute for Astronomy, University of Edinburgh, Royal Observatory, Blackford Hill, Edinburgh, EH9 3HJ, UK \label{inst2}
             \and
             INAF, Osservatorio Astronomico di Roma, Via Frascati 33, I– 00078 Monte Porzio Catone, Italy \label{inst9}
             \and
             School of Physics, HH Wills Physics Laboratory, University of Bristol, Tyndall Avenue, Bristol, BS8 1TL, UK\label{inst19}
             \and
             Department of Physics and Astronomy, York University, 4700 Keele Street, Toronto, ON M3J 1P3, Canada\label{inst13}
\and
Dipartimento di Fisica, Sapienza Universtià di Roma, Piazzale Aldo Moro 5, 00185 Rome\label{inst17}
\and
INAF - Osservatorio Astronomico di Trieste, Via Tiepolo 11, 34143 Trieste, Italy\label{inst12}
\and
IFPU - Institute for Fundamental Physics of the Universe, Via Beirut 2, 34014 Trieste, Italy \label{inst18}
\and
INAF – Osservatorio di Astrofisica e Scienza dello Spazio di Bologna, Via Gobetti, 93/3, 40129, Bologna, Italy\label{inst14}
\and
School of General Education, Shinshu University, 3-1-1 Asahi, Matsumoto, Nagano 390-8621, Japan\label{inst16}
\and
Research School of Astronomy and Astrophysics, Australian National University, Cotter Road, Weston Creek ACT 2611, Australia\label{inst10}
\and
Centre for Gravitational Astrophysics (CGA), Australian National University, Building 38 Science Road, Acton ACT 2601, Australia\label{inst11}
\and
Dipartimento di Fisica ed Astronomia (DIFA), Università di Bologna, Via Gobetti, 93/2, 40129 Bologna, Italy\label{inst15}}
 
      \abstract
   {We report the discovery of an extremely high-velocity outflow (EHVO) in the most luminous (\lbol $\sim$ 2.29 $\times$ 10$^{48}$ erg/s) QSO, SMSS J2157-3602, at z=4.692. Combined XSHOOTER and NIRES observations reveal that the EHVO reaches a maximum velocity of v$_\mathrm{max} \sim 0.13c$ and persists over rest-frame timescales of a few months to one year. SMSS J2157-3602 also exhibits one of the highest balnicity index discovered in an EHVO so far. In addition, the blueshifted CIV emission traces a high-velocity (v$\rm_{CIV}^{50}\sim$ 4660 km/s) outflow from the broad-line region. Thanks to an XMM-Newton observation, we also discover the X-ray weak nature of this QSO, which likely prevents the overionization of the innermost disk atmosphere and facilitates the efficient launch of the detected EHVO and BLR winds. 
The extraordinary luminosity of SMSS J2157-3602 and the extreme velocity of the EHVO make it a unique laboratory for testing AGN driven feedback under extreme conditions. Current uncertainties on the outflow’s location and column density strengthen the case for dedicated follow-up, which will be essential to assess the full feedback potential of this remarkable quasar.}


   \maketitle

\section{Introduction}

Outflows originating from the inner regions around Super-Massive Black Holes (SMBHs)
are detected in a substantial fraction of QSOs (up to $\sim$ 50\%) as blueshifted absorption lines in their UV and X-ray spectra (e.g. \citealt{Crenshaw03}, \citealt{Blustin05}, \citealt{Misawa07a}, \citealt{Bischetti22}),\citealt{Tombesi10}, \citealt{Matzeu23}).
The most dramatic nuclear outflows are represented by the ultra-fast (up to 0.2--0.5$c$) outflows (UFOs) discovered in X-rays through highly-blueshifted absorption features of He- and H-like Fe at $E>$ 7 keV, which originate at tens/hundreds gravitational radii from the SMBH (\citealt{Tombesi12}). Their large velocities imply large kinetic powers ($\dot{E}_{\rm  K,ufo}$), up to 10--20\% of the QSO bolometric luminosity, as $\dot{E}_{\rm  K,ufo}$ $\propto$ v$^3_{\rm  ufo}$ (\citealt{King2015}). 
UFOs may therefore be crucial to the study of feedback mechanisms, as they represent potentially the most energetic outflows due to their higher velocities, injecting an amount of energy into the surrounding ISM sufficient to significantly influence the evolution of the host galaxy. 

Absorption lines in rest-frame UV spectra are usually classified into three categories: broad absorption lines (BALs) with FWHM $\ge$ 2,000 \kms, narrow absorption lines (NALs) with FWHM $\le$ 500 \kms\ and mini-BALs (500 $\le$ FWHM $\le$ 2000 \kms). 
BALs have typically been identified within the velocity range of 5,000 to 25,000 {\kms} (\citealt{Weymann1991}). However, such features can reach velocities of $\sim$50,000-60,000 km s$^{-1}$ or more, prompting the introduction of a new term: extremely high velocity outflows (EHVOs; \citealt{RodriguezHidalgo11}).
Hereafter, the term "EHVO" denotes BAL outflows with velocities v $>$ 25,000 {\kms}.

Over the past decade, the number of sources detected with EHVOs has increased, expanding from individual quasars (e.g., \citealt{RodriguezHidalgo11}; \citealt{Rogerson16}; \citealt{Bruni19}) to hundreds of QSOs (\citealt{RodriguezHidalgo20}, hereafter RH2020, and Rodr\'iguez Hidalgo in prep).

Indeed, RH2020 recently conducted a pioneering survey of EHVOs in the general SDSS QSO population, analyzing 6743 QSOs and detecting 40 QSOs with EHVOs, which were found to be more prevalent at higher luminosities. These EHVOs were identified in the UV spectrum primarily through CIV and NV absorption at speeds between 10\% and 20\% of the speed of light, i.e., similar to X-ray UFOs. Specifically, the detection is based on the fact that SiIV absorption always has corresponding CIV absorption outflowing at similar speeds, with no cases reported so far in the literature of quasar outflows where SiIV absorption is present without corresponding CIV. Therefore, an absorption feature between Ly$\alpha$ and SiIV without corresponding CIV can be ascribed to a CIV EHVO.

Here, we report the discovery of an EHVO in the most luminous QSO known to date in the first 1.3 Gyr after the big bang at z = 4.692: SMSS J215728.21-360215.1 (hereafter SMSS J2157; \citealt{Wolf18}). 

Its discovery was enabled by the photometric and astrometric data from SkyMapper Southern Sky Survey (SMSS), Wide-field Infrared Survey Explorer (WISE), and Gaia, with an estimated bolometric luminosity of 1.6 $\times$ 10$^{48}$ erg/s, inferred from the monochromatic luminosity at 3000 $\AA$ (\citealt{Onken20}). An anisotropy-corrected bolometric luminosity based on the spectral energy distribution (SED) is presented in \cite{Lai2023} and and SED-based estimate derived in this work is discussed in Sect. \ref{sec:SED}. SMSS J2157 was not detected in any large radio survey (National Radio Astronomy Observatory Very Large Array Sky Survey (NVSS) f$\rm_{1.4GHz}$<2.5 mJy (\citealt{Condon1998}), Sydney University Molonglo Sky Survey (SUMSS) f$\rm_{843MHz}$<5.0 mJy (\citealt{Bock1999})), therefore it is classified as radio-quiet QSO (\citealt{Wolf18}).

It hosts a SMBH with an estimated MgII-based BH mass M$\rm_{BH}\sim$3.4$\times$ 10$^{10}$ M$_{\sun}$, positioning it among the largest black holes known to date, and an Eddington ratio \edd$\sim$0.4 (\citealt{Onken20}; see also \cite{Lai2023} for accretion disc-based BH mass estimate and corresponding Eddington ratio).
Recent observations using the XSHOOTER instrument (\citealt{Vernet2011}) at the Very Large Telescope (VLT) and the Near-Infrared Echellette Spectrometer (NIRES) instrument (\citealt{Wilson2004}) at Keck Observatory have suggested the presence of a potential EHVO in SMSS J2157 spectra. In this paper, we analyze the absorption feature observed in the rest-wavelength range 1350–1400 $\AA$, confirming its EHVO nature with a maximum velocity of $\sim$39,500 km/s. In Sect. \ref{sec:obs}, we present the spectroscopic observations and the reconstruction of the UV spectrum. In Sect. \ref{sec:results}, we characterize the EHVO detected in SMSS J2157. Sect. \ref{sec:SED} provides results from spectral and photometric analysis and Sect. \ref{sec:column_density} present the total column density derivation. In Sect. \ref{sec:xray}, we describe the XMM-Newton observations, including data reduction and analysis, and discuss the X-ray properties of SMSS J2157. In Sect. \ref{sec:kinematics}, we analyze the kinematics and energetics of the EHVO in SMSS J2157, highlighting the associated large uncertainties. Finally, in Sect. \ref{sec:conclusion}, we summarize our findings. In this work, we assume that $\Omega\rm_m$ = 0.27, $\Omega_\Lambda$ = 0.73, and $H_0$ = 70 km s$^{1}$ Mpc$^{-1}$ (\citealt{Komatsu2011}).


\section{Observations and data reduction}\label{sec:obs}

\subsection{Spectral datasets}

We analyzed spectroscopic observations obtained with the XSHOOTER instrument (\citealt{Vernet2011}) at the Very Large Telescope (VLT). 
Details of the observations are listed in Table \ref{table:1}.
Specifically, we used XSHOOTER data (Program ID: 0103.B-0949(A)) with SMSS J2157 observed in 2019 June 03 and 2019 July 08. The data reduction of the XSHOOTER data was performed using the XSHOOTER pipeline version within the ESO-Reflex environment (version 2.11.5; \citealt{Modigliani2010}). Science spectra for each arm were individually reduced, adopting the nodding mode setting in the pipeline. For flux calibration, standard stars were observed and reduced using the same calibration data as the science frames. Telluric line correction was conducted with molecfit (\citealt{Smette2015}). Finally, we applied the barycentric velocity correction for each spectrum. For simplicity, we refer to these data as XSHOO-1 and XSHOO-2 spectrum, respectively. We also included in our analysis the dataset presented by \citet{Onken20}, which combines Keck/NIRES spectra from June 2018 with VLT/XSHOOTER data from October 2019. Hereafter, we refer to this dataset as NIRES-XSHOO. A detailed description of the spectroscopic observations and data reduction procedures can be found in \cite{Onken20}.

We rescaled XSHOO-1, XSHOO-2 and NIRES-XSHOO spectra to the J-band VHS Data Release 6 (J$\rm_{Vega}$=15.65 $\pm$ 0.01, \citealt{Wolf18}) and SkyMapper z-band magnitudes (z$\rm_{AB}$=17.11 $\pm$ 0.02, \citealt{Wolf18}).

\begin{figure*}
    \centering
    \includegraphics[width=18cm]{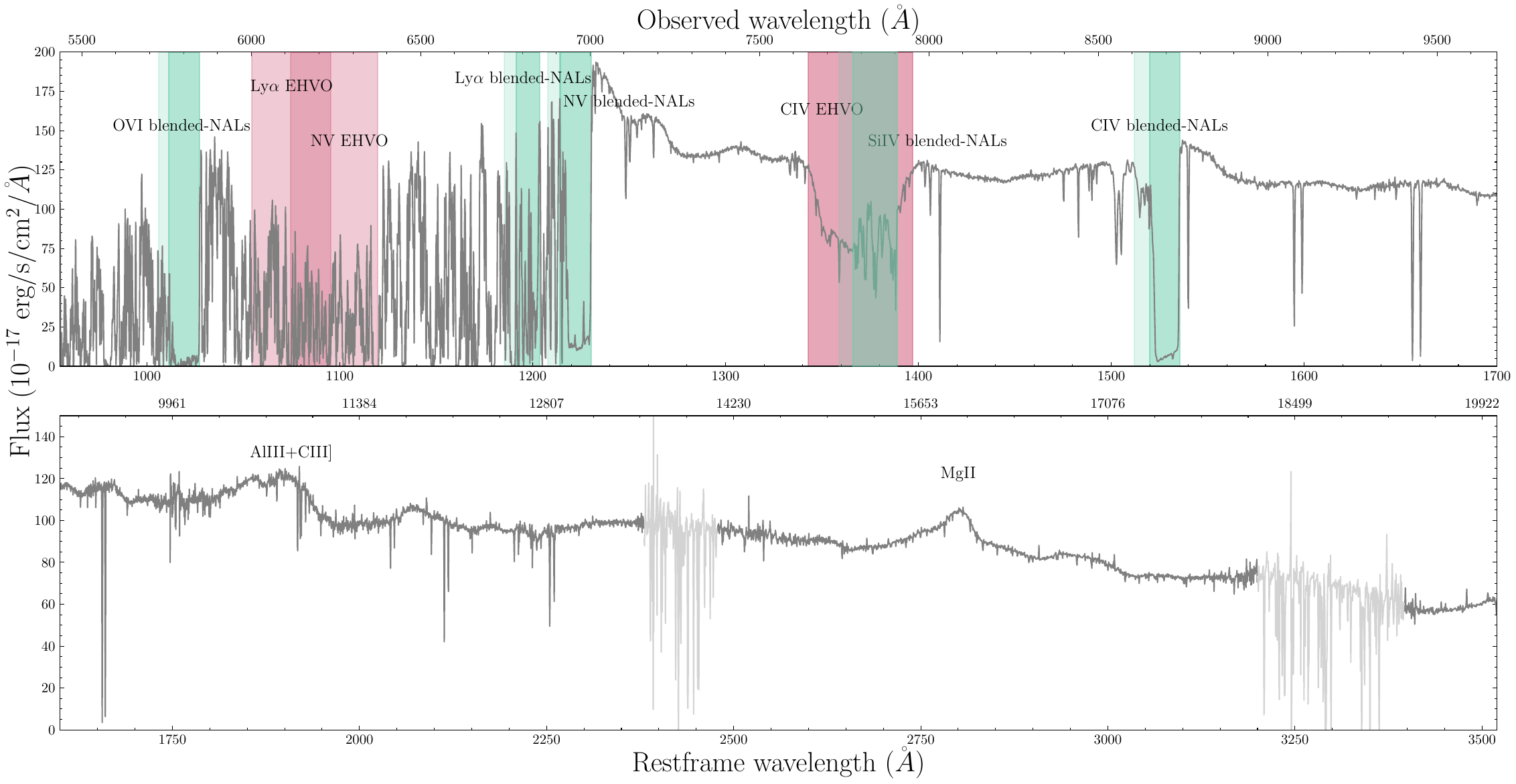}
    \caption{Final coadded spectrum of SMSS J2157, with the CIV, NV and Ly$\alpha$ EHVO highlighted in light red. The blended NALs systems are indicated in aquamarine. The AlIII+CIII] and MgII emission lines are also marked. The regions affected by telluric absorption are marked as light grey color.}
    \label{fig:spectrum_final}
\end{figure*}

No significant variability was observed between the June and July observations of XSHOO-1 and XSHOO-2 spectra, we have then averaged them and rebinned the final coadded 1D spectrum to match the NIRES-XSHOO spectrum bin size (50 {\kms}). Since no further variability was detected between this final coadded 1D spectrum and the NIRES-XSHOO spectrum, both spectra were also averaged. Additionally, public light curve data from the NASA/ATLAS survey in the o-band (orange filter, $\sim$560–820 nm) show only minor variability over a four-year period (2017–2021), supporting our findings. The final spectrum is shown in Fig. \ref{fig:spectrum_final}.

\begin{table*}
\centering
\begin{threeparttable}
\caption{Journal of observations.}\label{table:1}             
\begin{tabular}{lllcccc}       
\hline\hline               
     OBS ID &    Instrument & Obs. Date & P.I. &  Exp. Time (s) & Median seeing (arcsec) & Slit width (arcsec)\\
         \hline                       
      XSHOO-1&  VLT/XSHOOTER &  2019 June 03 &Zappacosta & 1780, 1680, 1800*   & 0.68 & 1, 0.9, 0.9* \\
    XSHOO-2 &VLT/XSHOOTER &  2019 July 08 &  Zappacosta  & 3560, 3360, 3600*  & 0.68 &  1, 0.9, 0.9* \\
  
\end{tabular}
\begin{tablenotes}
\footnotesize
\item[*] The exposure times and slit width refer to the three arms UVB, VIS, and NIR of the XSHOOTER spectrograph.
\end{tablenotes}

\end{threeparttable}
\end{table*}

\subsection{Reconstruction of the rest-frame UV spectrum}
\label{sec:norm}

As shown in Fig. \ref{fig:spectrum_final}, several low-velocity (v $\leq$ 25,000 km/s; shades of cyan) BAL features are present, with matching velocity ranges blueward of CIV, SiIV, NV, and OVI. The absence of an absorption feature blueward of MgII indicates that SMSS J2157 is a high-ionization BAL quasar, in contrast to low-ionization BAL quasars, which also show absorption from low-ionization species such as MgII.

However, as described in Sect. \ref{sec:bal} and shown in Fig.\ref{fig:CIV_blended}, although the CIV trough satisfies the classical observational definition of a BAL (e.g., \citealt{Weymann1991}), our line profile modeling reveals that it results from the superposition of multiple narrow absorption components (FWHM $<$ 500 km s$^{-1}$), indicating that this is a blended-NAL system.

These absorption features are superimposed on their respective emission lines. The trough blueward of SiIV is significantly broader in velocity than the absorption features blueward of CIV, suggesting that part of this feature may be attributed to an EHVO of the CIV ion (RH2020). This absorption is likely accompanied by an EHVO of NV and possibly Ly$\alpha$ at a similar velocity, which fall within the Ly$\alpha$ forest.
To properly account for the complex blend of SiIV emission line, where both the SiIV blended NALs and the CIV EHVO are present, as well as the corresponding NV and Ly$\alpha$ EHVO in the Ly$\alpha$ forest region, we adopt a two-step approach.

\begin{figure}
    \centering
    \includegraphics[width=9cm]{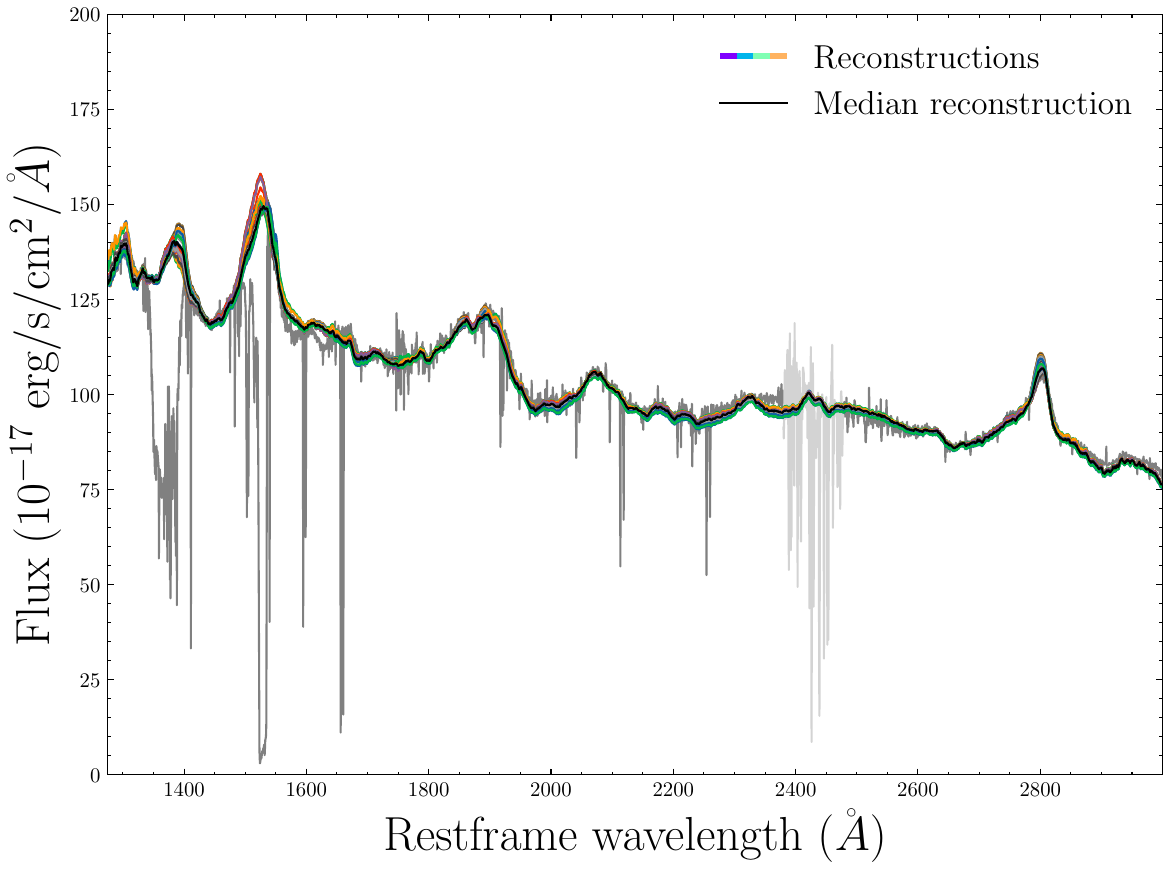} 
    \caption{Median spectral reconstruction (black) created from the reconstructions with different parameters used in the spectral fitting. The original spectrum is plotted in grey. The regions affected by telluric absorption are marked as light grey color.}
    \label{fig:recon}
\end{figure}

We first derive the best-fit model of the continuum, using a custom Python-based code (e.g. \citealt{Vietri22}) to fit simultaneously the continuum with a power-law and the principal emission lines as Ly$\alpha$,  NV$\lambda$1240, SiV$\lambda$1398, OIV$\lambda$1402, CIV$\lambda$1549, HeII$\lambda$1640, OIII]$\lambda$1663, AlIII$\lambda$1857, SiIII$\lambda$1887 and CIII]$\lambda$1909, using Gaussian components while masking all narrow and broad absorption lines. The model fitting is performed in the spectral region at 1210-2000 $\AA$, to avoid that intervening absorption on the blue side of Ly$\alpha$ affect the fit. We therefore extrapolate the continuum to the blue side of the Ly$\alpha$ from the best-fit model.
We then used a reconstruction model to provide a best estimate of the emission lines and to avoid underestimating the strength of the absorption features.
The reconstruction was generated via the scheme developed for \citet{Rankine20} which is based on an Independent Component Analysis (ICA; \citealt{Hojen-Sorenson2002Mean-FieldAnalysis}; \citealt{Opper2005ExpectationInference}; \citealt{Allen2013ClassificationAnalysis}) of SDSS quasars. The ICA components are linearly combined to reproduce the intrinsic emission and is particularly useful when much of the emission has been absorbed. The ICA components were generated from and for reconstructing low-S/N SDSS spectra; however, the data used here are of higher resolution (c.f. R$\sim$2000 for SDSS, R$\sim$5400, 8900, 5600 for UVB, VIS and NIR arms of XSHOOTER, respectively) and higher S/N. In order to reconstruct the spectrum, we first re-bin the spectrum onto the $\Delta \log\lambda=0.0001$ wavelength grid of SDSS \citep[using {\sc SpectRes};][]{spectres}. The reconstruction scheme can then be applied to the spectrum to produce a reconstruction covering 1275--3000\,{\AA}. Part of the routine masks bad pixels and narrow absorption features that are $N$-$\sigma$ below the continuum level (where $\sigma$ is the noise array). Since the routine was fine-tuned for SDSS-level S/N and resolution, we repeated the spectral fitting with a grid of $N$ values: $N=1/8, 1/6, 1/4, 1/2, 1, 2, 3$, and we also degraded the S/N of the spectrum by multiplying the noise array by a factor of 8, 6, 4, 2 and 1. The reconstructed spectra created with all combinations of these parameters are presented in Fig.~\ref{fig:recon} alongside the median reconstruction. On first inspection the reconstructions appear to differ significantly at the CIV and SiIV emission; however, the majority of the individual reconstructions are similar to the median reconstruction and only a few infer stronger emission lines. The median reconstruction was then adopted to normalize the spectrum between 1275--3000\/{\AA} along with the best-fit model continuum blueward of the Ly$\alpha$ line as shown in Fig. \ref{fig:CIV_EHVO_bestfit}.

From the median reconstruction, we perform a direct line integration to derive the CIV emission line equivalent width, obtaining EW(CIV)$\sim$17$\pm$2 $\AA$, placing it marginally within the weak emission line regime as defined by \citet{Chen2024}. We also calculate the 50th percentile velocity, v$\rm_{CIV}^{50}\sim$ 4660$\pm$200 km/s \footnote{We adopted the convention of using a positive sign for blue-shifted line velocities., with respect to the MgII-based redshift (\citealt{Onken20})}, indicating a strong outflow originating from the broad line region (BLR). These values of EW and velocity shift of the CIV emission line are consistent with the properties found for the EHVO sample analyzed in \citealt{RodriguezHidalgo22}. EHVOs are characterized by low EW(CIV) and significant blueshifts in the CIV emission line, exhibiting values that are more extreme than the average observed in both non-BAL and BAL QSOs (\citealt{RodriguezHidalgo22}).

\subsection{Blended SiIV NAL removal from CIV EHVO region}\label{sec:bal}

As mentioned above, the CIV EHVO falls in the same location as the SiIV blended NAL at lower velocity. To not overestimate the amount of CIV EHVO, we proceeded to fit and remove the SiIV from this region in an iterative process. 

We assumed that the normalized observed flux at a given velocity \(I(v)\) is described by:
\begin{equation}
    I(v) = [1 - C_f] + C_f e^{-\tau(v)}
\end{equation}
The parameter \(C_f\) is the coverage fraction ($0 < C_f < 1$; \citealt{Hamann99}), and $\tau(v)$ is the optical depth, which we assumed to follow a Gaussian profile characterized by the central optical depth ($\tau_0$), centroid velocity ($\mu$), and Doppler parameter ($b$). 

Initially, we plotted the SiIV doublet\footnote{Each analyzed doublet was constrained to have the correct separation and a 2:1 optical depth ratio between the short- and long-wavelength components.} visually adjusting the component widths and centroid velocities to match the observed profiles. We then fitted the corresponding CIV blended NAL region (see Fig. \ref{fig:CIV_blended} in Appendix \ref{app:app}), maintaining the same widths and centroid velocities derived from the SiIV region, and allowing only the optical depths to vary. The initial fit indicated the need for additional, weaker NAL components. These were iteratively added until a satisfactory fit was achieved. We observed the CIV blended NAL system to be saturated at about 2\% of the continuum flux level, which is a sign of $C_f$ lower than 1. The best-fit solution yielded a constant $C_f = 0.98$ value for this region.  
The final component widths and centroid velocities were then adopted to model the SiIV NAL doublets as described in the next section.

\section{Results from spectral and photometric analysis}\label{sec:results}

\subsection{Analysis of the EHVO}\label{sec:analys}

To model the CIV EHVO absorption, which is blended with lower-velocity SiIV NALs, we performed a simultaneous fit of both components. For the SiIV blended NALs, we kept fixed Doppler parameters and centroid velocities, as derived in Sect. \ref{sec:bal}, and left free to vary the optical depths. 

\begin{figure*}
    \centering
    \includegraphics[width=\linewidth]{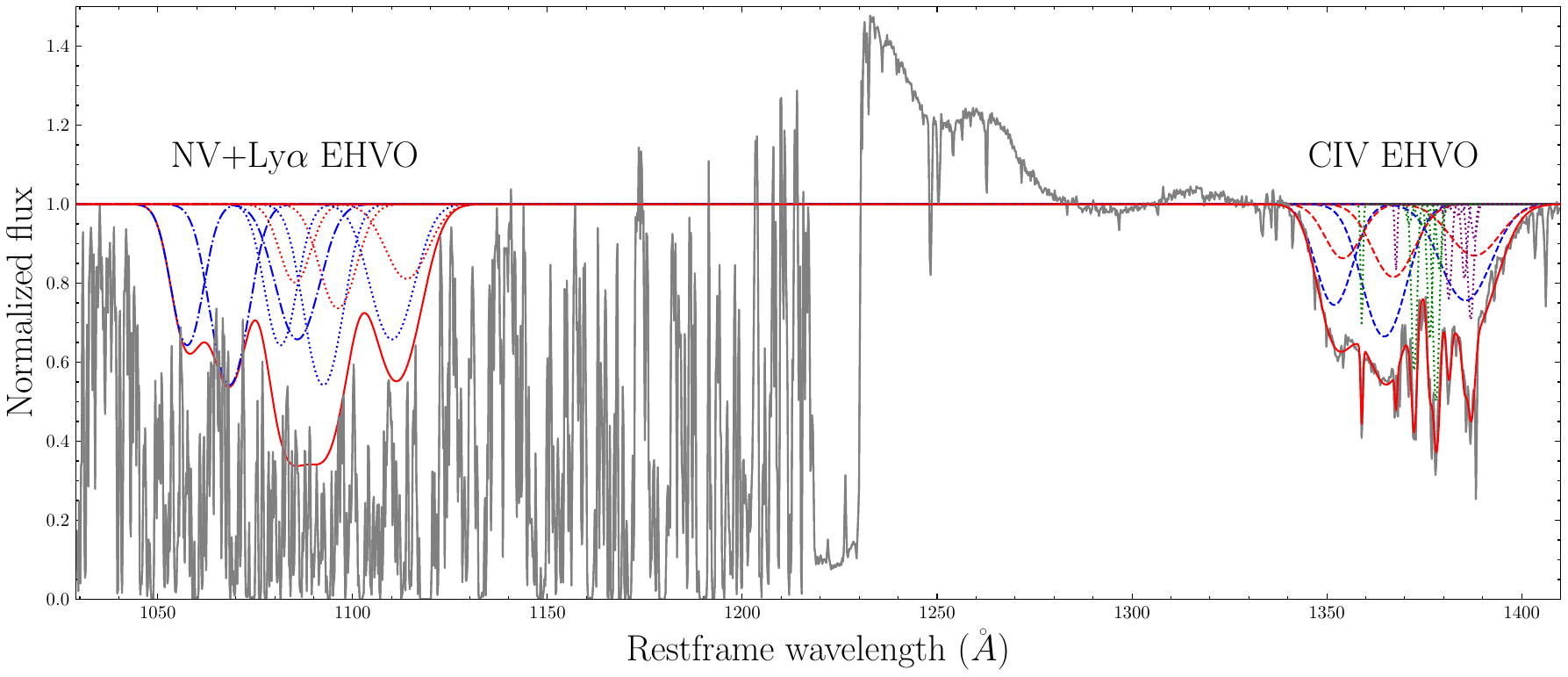}
  
  \caption{Normalized, coadded spectrum of SMSS J2157, overlaid with the best-fit model (red curve) for the EHVO features of the CIV doublet (three blue-red pairs of dashed Gaussian components for modeling 1548,1550 $\AA$ lines respectively), NV doublet (three blue-red pairs of dotted Gaussian curves for modeling 1238,1242 $\AA$ lines respectively), Ly$\alpha$ (three blue dash-dotted Gaussian curves), and SiIV doublet blended NALs (seven green-purple pairs of dotted Gaussian components for modeling 1393,1402 $\AA$ lines, respectively). The spectrum was normalized to the reconstructed model at $\lambda>$1275 \AA\ and to the extrapolated continuum $\lambda<$1275 \AA.}
    \label{fig:CIV_EHVO_bestfit}
\end{figure*}

We model the CIV EHVO doublet with a Gaussian profile for each component, as described in Sect.~\ref{sec:bal}. Up to three doublets are used to model the CIV EHVO absorption. The central optical depth, centroid velocity, and Doppler parameter of each component are free to vary. 
The covering fraction $C_f$ was systematically explored by evaluating values between 0.1 and 1. For each fixed $C_f$, we performed 100 fits with randomized initial parameters for all Gaussian components.
The global best-fit solution is obtained for $C_f = 1$, which minimizes both $\chi^2$ and the Bayesian Information Criterion (BIC), and is shown in Fig.~\ref{fig:CIV_EHVO_bestfit}. We therefore adopt the $C_f$=1 solution as our fiducial best-fit model. Solutions with very low $C_f$ ($\approx 0.1$--0.41), corresponding to a quasi-saturated regime, are strongly disfavored by the fit statistics ($\Delta$BIC $\gg 10$), whereas solutions with $C_f \ge 0.44$ are statistically equivalent ($\Delta$BIC $< 10$) and cannot be formally excluded. 
This range of acceptable $C_f$ values allows us to estimate a conservative range for the hydrogen column density $N_{\rm H}$, by considering $N_{\rm H}$ derived for $C_f^{\rm min} = 0.44$ and $C_f^{\rm max} = 1$ (see Sect.~\ref{sec:column_density}).

We then use the CIV EHVO best-fit model as a template to fit the NV doublet and Ly$\alpha$ EHVO troughs, which are blended with the Ly$\alpha$ forest. In this fit, only the optical depths are allowed to vary, while the Doppler parameters and centroid velocities are kept fixed. Since the NV and Ly$\alpha$ absorption features fall within the Ly$\alpha$ forest, we treat our measurements in this region as upper limits. Figure \ref{fig:CIV_EHVO_bestfit} also shows the best-fit model for the NV doublets and Ly$\alpha$ EHVO.

From the best-fit model of the CIV EHVO, we measured the Balnicity index (BI; \citealt{Weymann1991}) to characterize the CIV EHVO absorption by adopting 25,000 and 60,000 km/s as minimum and maximum velocity integration limits:

\begin{equation}
\mathrm{BI} = - \int_{60{,}000}^{25{,}000} \left[1 - \frac{f(\mathrm{v})}{0.9} \right] C \, d\mathrm{v}
\label{eq:1}
\end{equation}

where we adopted as \(f(\mathrm{v})\) the CIV EHVO best-fit model as a function of the velocity v, and \(C\) is a constant set to unity if the spectrum is at least 10 per cent below the continuum model for velocity widths of at least 1000 {\kms} and zero otherwise.

We calculated the CIV BI$\rm_{EHVO}^{CIV}$ = 2200 km s$^{-1}$, placing it in the top $\sim$20\% of the BI distribution for BAL quasars at lower redshifts (\citealt{Gibson09a}) and among the largest values discovered so far for EHVOs with velocities exceeding 35,000 km s$^{-1}$ (see RH2020). We derived a minimum velocity for the CIV EHVO of v$\rm_{min}\sim 31170$ km s$^{-1}$ and a maximum velocity of v$\rm_{max}\sim 42350$ km s$^{-1}$. Velocities are given relative to the longer-wavelength component of the CIV doublet, assuming the relativistic Doppler effect. 
We also derived an upper limit for BI of the NV doublet and Ly$\alpha$ EHVO, i.e. BI$\rm_{EHVO}^{NV}$$\le$4260 km s$^{-1}$ and BI$\rm_{EHVO}^{Ly\alpha}\le$2800 km s$^{-1}$, respectively.

 \subsection{Spectral Energy Distribution}\label{sec:SED}

We model the SED spanning from the X-ray to MIR-infrared wavelengths.
We used publicly available photometric data in the YJKs bands from the VISTA Hemisphere Survey (\citealt{McMahon2013}), in the H band from the Two Micron All-Sky Survey (\citealt{Skrutskie2006}), and W1-W4 bands from the Wide-field Infrared Survey Explorer (\citealt{Wright2010}). Comprehensive details about the photometry can be found in the discovery paper (\citealt{Wolf18}).  

Additionally, we used observations obtained with the Rapid Eye Mount (REM) telescope at the La Silla Observatory in November 7, 2019 (PI: V. Testa), using the ROS2 visible photometric channel. 

A series of 9 images were acquired with the SDSS-$griz$ filters, and each exposure lasted $240$~s. All the images were processed using the jitter script from the \texttt{eclipse} package (\citealt{Devillard1997}). This script aligns and stacks series of images to create an average frame for each sequence, while also performing sky subtraction. The magnitudes measured for the optical $griz$ filters have been calibrated against several field stars selected from the APASS catalogue (DR9, \citealt{Henden2016}), by performing differential photometry using an aperture of 8 pixels (corresponding to $\sim 4.6$~arcsec). Table \ref{tab:my_label} lists the REM AB magnitudes, corrected for the galactic extinction using values reported in \cite{Schlafly2011}.
We also used X-ray data point obtained with XMM-Newton and presented in Sect. \ref{sec:xray}.

\begin{figure}
    \centering
    \includegraphics[width=9cm]{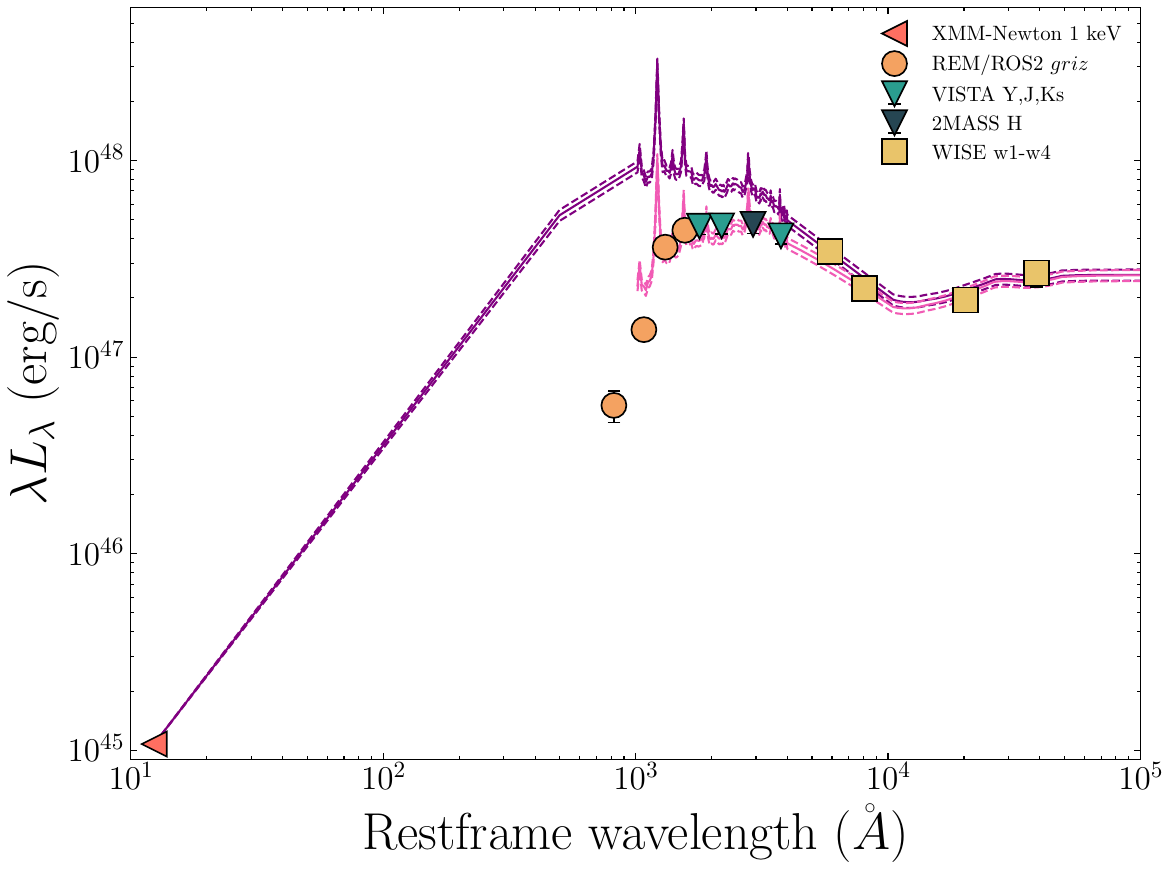}
    \caption{Best-fit de-reddened and reddened SEDs with 1$\sigma$ errors for SMSS J2157 (purple and pink lines, respectively), as derived in Sect. \ref{sec:SED}. The 1 keV luminosity is shown as a left-pointing coral red triangle; REM/ROS2 $g,r,i,z$ data are represented by orange circles; Y, J, and Ks points from VISTA are shown as down-pointing green triangles; the H point from 2MASS is represented by a down-pointing blue triangle; and golden yellow squares represent the W1-W4 WISE channels. The uncertainties of data points are shown but are smaller than the symbol size.}
    \label{fig:SED_best}
\end{figure}

We performed SED fitting for the luminosity points at $\lambda \geq 1216$ \AA. To avoid having the fit driven by points with very low uncertainties, we added an error of 0.1 mag in quadrature for each data point's uncertainty (e.g., \citealt{Boquien2019}). To model the SED, we used three templates of optically selected Type-1 QSOs, distinguished by increasing infrared-to-optical flux ratios, i.e. BQSO1, QSO1, TQSO1, taken from the SWIRE template library \cite{Polletta2007}. 
Given the extreme luminosities of the source, the host emission can be considered negligible (e.g., \citealt{Shen2011}).
For each template, we maximized the likelihood using \(emcee\) \citep{Foreman-Mackey2019}:

\begin{equation}
    \log L(K, E(B-V)) = -\frac{1}{2}\sum_{i} \left( \frac{y_{i}- K f_{i}10^{-0.4A_{\lambda}}}{\sigma_{i}}\right)^{2} + \log(2\pi \sigma_{i}^{2})
\end{equation}

where $y_{i}$ represents the photometric data points at the $i$-th filter, and $f_{i}$ denotes the flux obtained by convolving the SED template with the $i$-th filter. $K$ is the normalization factor of the template, and $A_{\lambda} = K_{\lambda}(E(B-V))$ is the dust-reddening law from \cite{Prevot1984}, accounting for possible contributions from dust extinction. For both $K$ and $E(B-V)$, we assumed flat, non-negative priors. Since the BQSO1 template provided the best fit, yielding the highest likelihood value, we report quantities computed with it. We obtained a color excess $E(B-V)$= $0.07 \pm 0.005$, where uncertainties are quoted as the 16th and 84th percentiles. The best-fit de-reddened (purple curve) and reddened (pink curve) SEDs are shown in Fig. \ref{fig:SED_best}.

The bolometric luminosity was computed by integrating the best-fit SED from 1 \textmu m to 1 keV, resulting in Log($L\rm_{Bol}$/erg s$^{-1}$) = 48.36. Hereafter, we will adopt this value of $L_{\rm Bol}$ in our analysis. Since the employed templates only extend up to 900 \AA, we extrapolated the EUV region of the SED using the double power-law recipe from \cite{Lusso2010} and \cite{Saccheo2023}, i.e., $\lambda L_{\lambda} \propto \lambda^{0.8}$ down to $\lambda = 500$ \AA, plus a power law with a free-to-vary index to connect the 1 keV luminosity with the 500 \AA\ one.

\begin{table}[!h]
    \caption{SMSS J2157 optical magnitudes obtained with REM.}
    \centering
    \begin{tabular}{c|c}
       \hline
       ROS2 filter  &  Magnitude\\
       \hline
        $g$ & $\ge$ 19.86\\
        $r$ & 18.62$\pm$ 0.05 \\
        $i$ & 17.37$\pm$ 0.03 \\
        $z$ &  16.96$\pm$ 0.03 \\ 
       \hline
       \end{tabular}
    \label{tab:my_label}
\end{table}

\subsection{Ionic column densities and photoionization solution}\label{sec:column_density}

Using the Gaussian profiles of the CIV and NV doublets and Ly$\alpha$ corresponding to the minimum and maximum acceptable covering fractions ($C_f^{min}=$0.44 and  $C_f^{max}$=1), as derived in Sect. \ref{sec:analys}, we calculated the ionic column density \(N_\mathrm{ion}\) (\citealt{Arav01}) as follows:

\begin{equation}\label{eq:nion}
N{\rm_{ion}}=\frac{3.7679 \times 10^{14}}{\lambda f} \int \rm \tau(v)dv 
\end{equation}

where $\lambda$ is the laboratory wavelength and \(f\) is the oscillator strength corresponding to a specific ion. 
We used the numerical code Cloudy (version C23, \citealt{Chatzikos2023}) to compute the fraction of ionic species by varying the ionization parameter \(U\), assuming gas in photoionisation equilibrium and solar abundances (\citealt{Lodders2003}), and the broad-band SED of SMSS J2157, described in Sect. \ref{sec:SED} (see Fig. \ref{fig:SED_best}).

We found \Nion\ = 3.3$\times$10$^{15}$ cm$^{-2}$ (1.3$\times$10$^{16}$ cm$^{-2}$), \Nion\ = 9.3$\times$10$^{15}$ cm$^{-2}$ (3.7$\times$10$^{16}$ cm$^{-2}$) and \Nion\ = 3.6$\times$10$^{15}$ cm$^{-2}$ (1.5$\times$10$^{16}$ cm$^{-2}$) for CIV, NV and Ly$\alpha$ ions, respectively, for $C_f^{max}$ ($C_f^{min}$).

\begin{figure}
    \centering
    \includegraphics[width=\linewidth]{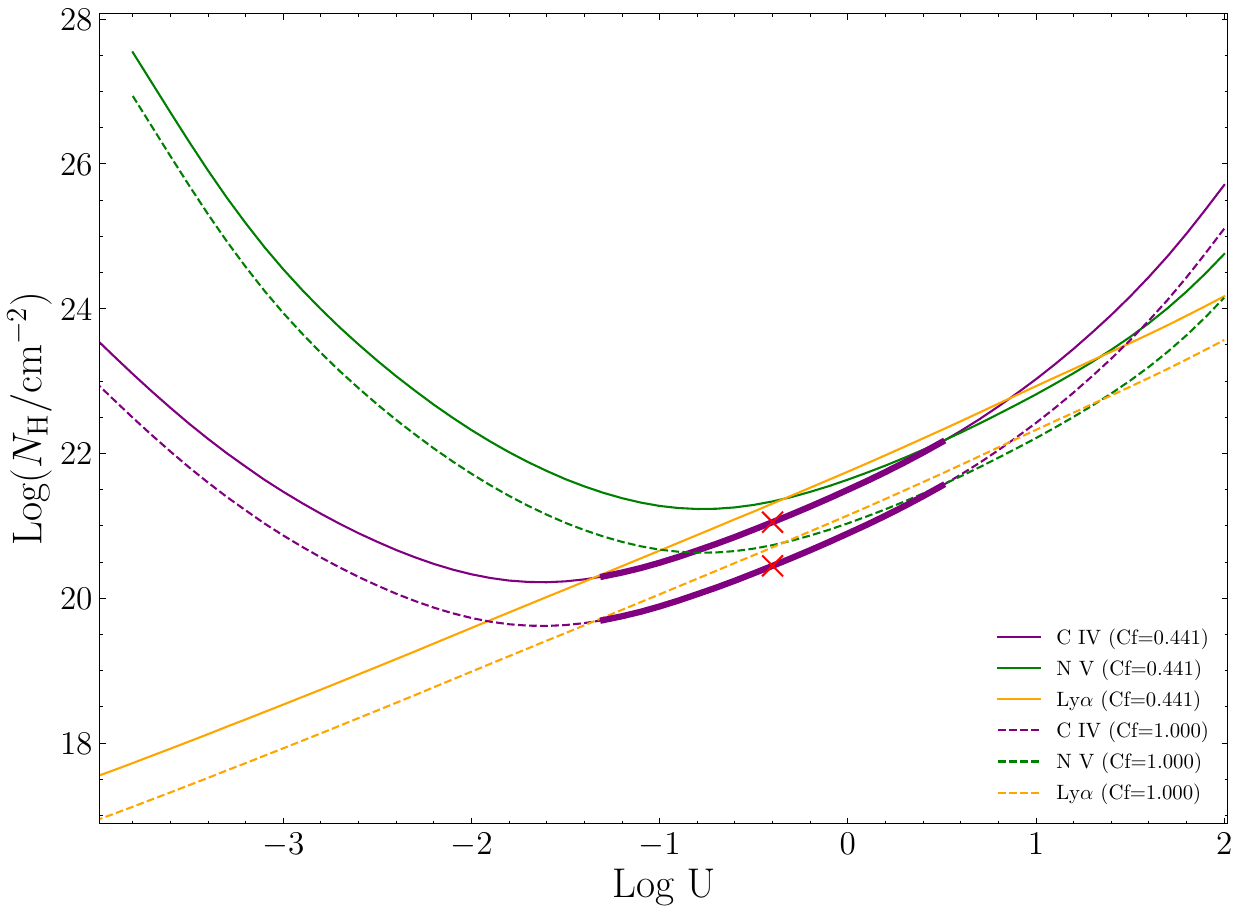}
   \caption{Theoretical values of the ionization parameter $U$ and the computed column density \NH\ from \Nion\ as derived from Eq. \ref{eq:nion} (see Sect. \ref{sec:column_density}). Solid and dashed lines correspond to the values of \NH\ derived using $C_f^{\rm min}$ and $C_f^{\rm max}$, respectively. The purple curves show the \NH\ constraint from CIV, while the green and orange curves represent the upper limits from NV and Ly$\alpha$, respectively. Bold lines indicate the allowed solutions for \NH\ and $U$, and the red crosses mark the median values of the solutions adopted in Eq.~\ref{eq:ekin}.}
    \label{fig:NH_U}
\end{figure}

Using the fraction of ionic species and ionic column densities calculated from Eq. \ref{eq:nion}, we derived the total hydrogen column density \NH\ as a function of \(U\), for both $C_f^{min}$ and $C_f^{max}$. Fig. \ref{fig:NH_U} shows the \NH\ curves as a function of \(U\), along with the possible solutions which are highlighted in bold lines, considering \NH\ curves of Ly$\alpha$ and NV as upper limits. From this analysis, we found \NH\ =[4.9$\times$10$^{19}$–3.5$\times$10$^{21}$] cm$^{-2}$ for $C_f^{max}$ and \NH\ =  [2$\times$10$^{20}$–1.4$\times$10$^{22}$] cm$^{-2}$ for $C_f^{min}$. For a conservative estimate, we adopt the median values, \NHCfmax=2.8$\times$10$^{20}$ cm$^{-2}$ and \NHCfmin= 1.1$\times$10$^{21}$ cm$^{-2}$ for $C_f^{max}$ and $C_f^{min}$, respectively, with \(U\) = 0.39, providing conservative bounds on \NH\ while accounting for the mild degeneracy with $C_f$. Relativistic effects were also considered (see Eq. 14 in \citealt{Luminari2024}).

\section{X-ray properties of SMSS J2157}\label{sec:xray}

\subsection{XMM-Newton data reduction}

SMSS J2157 has been observed in the X-rays with one exposure of 78.5~ks (Obs ID: 0844020101; PI. L. Zappacosta) with the observatory XMM-Newton, on 24–25 October 2019. The observation was reduced with the Science Analysis System (SAS) version~19.1.0 following the standard science threads. Time intervals characterized by highly variable high-energy background flares in the EPIC pn and MOS cameras were selected and removed. Specifically, we considered  good time intervals for scientific analysis only those having pn, MOS1 and MOS2 count-rates  $\leq0.4$~counts/s (in the $10-12$~keV band),  $\leq0.17$~counts/s ($>10$~keV) and $\leq0.24$~counts/s ($>10$~keV), respectively. We performed for each camera source spectral extraction on circular regions centered on the QSO optical position. The background region on the pn camera was chosen as a rectangular region around the source with roughly the same position angle as the observation and with short and long sides of $\sim2.3$ and $\sim3.7$~arcmin, while on the MOS cameras was taken to be a circular region centered on the QSO position with radius $\sim2.5$~arcmin. The extraction of the background spectra was performed on these regions excluding  circular regions of radius 40-50~arcsec (according to the source flux) centered on the position of all X-ray point-like sources (including the target QSO). The pn and MOS spectra were binned following the \citet{kaastra2016} optimization scheme which ensures high accuracy in recovering the true source spectral parameters regardless of the analysis energy interval adopted (\citealt{Zappacosta2023}). 

\subsection{X-ray spectral fitting results}

\begin{figure*}
    \begin{center}
    \includegraphics[width=0.48\textwidth]{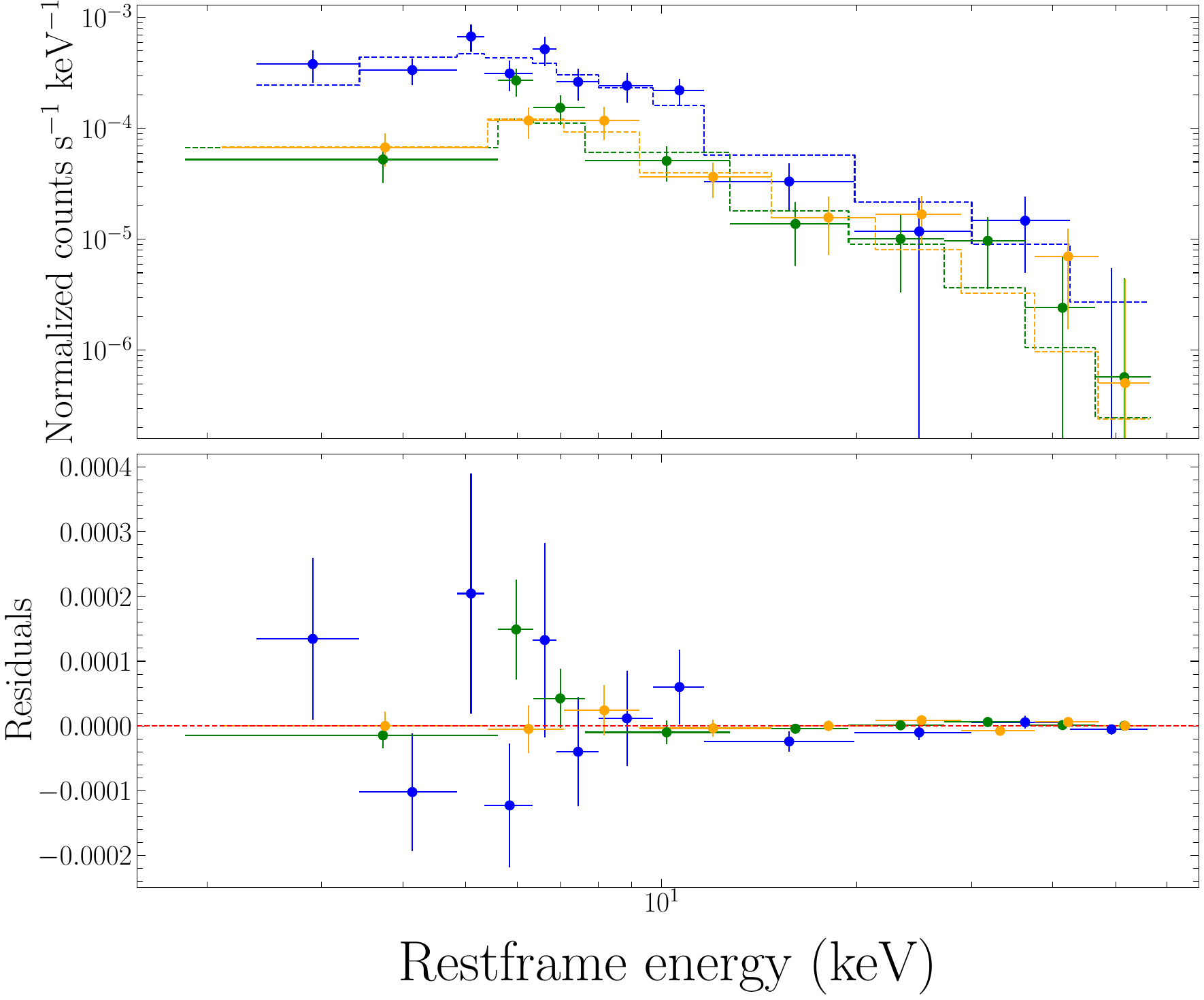}
    \includegraphics[width=0.48\textwidth]{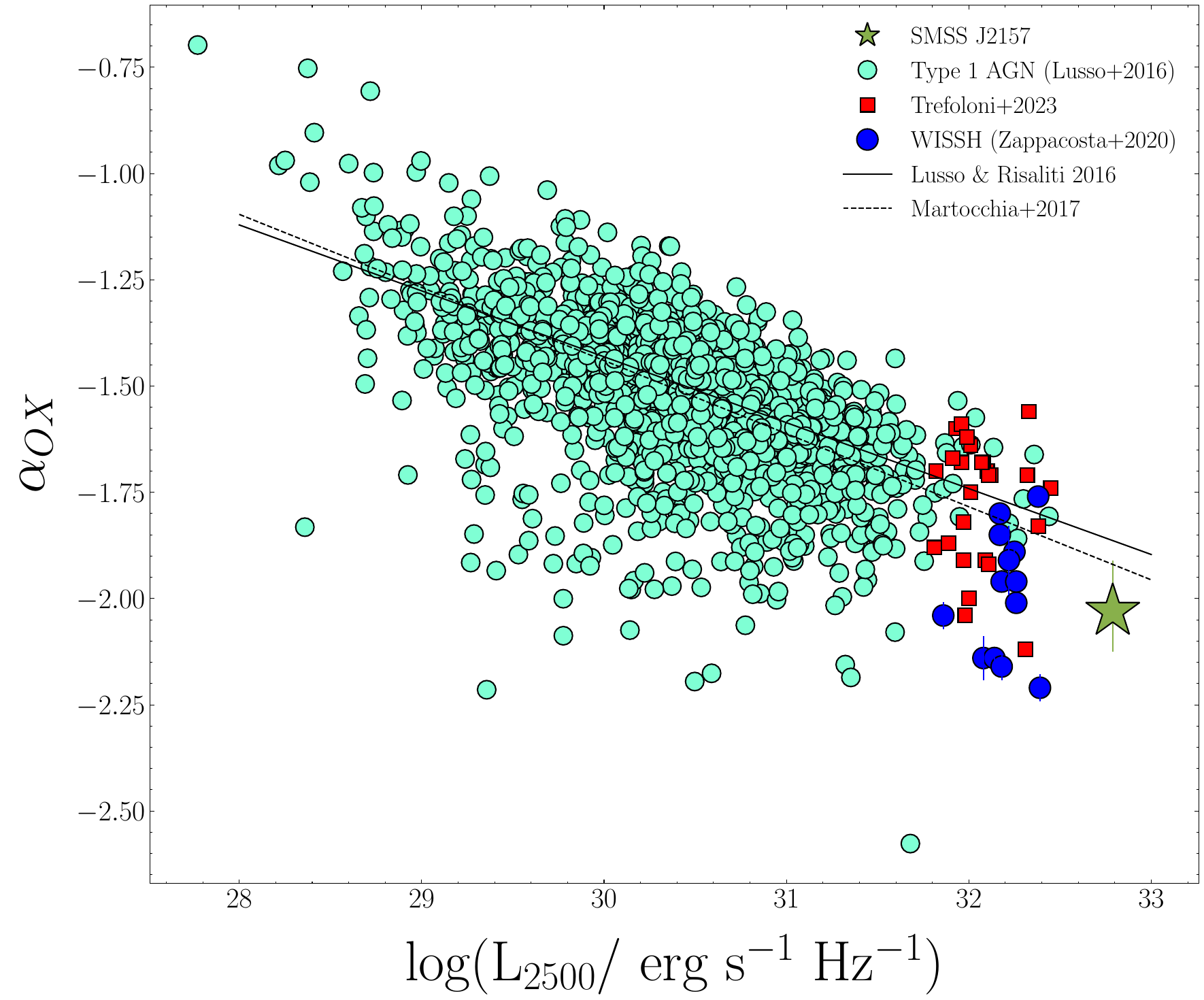}
    \end{center}
    \caption{Left panel shows in the upper plot the X-ray EPIC spectra (data) and relative best-fit power-law models modified by intrinsic absorption (dashed lines) and in the lower plot the residuals, i.e. data minus best-fit models. Blue, green and orange refer to the pn, MOS1 and MOS2 detectors, respectively. Energies are reported at the rest-frame. Right panel shows $\alpha_{\rm ox}$ vs $L_{2500}$. SMSS J2157 is reported as green star, while green, red and blue colors indicate a compilation of Type~1 AGN \citep{lusso2016}, bright UV-selected quasars  \citep{nardini2019} and optically/IR selected bright QSOs from the WISSH sample \citep[][]{zappacosta2020}, respectively. Solid and dashed black lines report the relations inferred by \citet{lusso2016} and \citet{martocchia2017}.}
    \label{xrayspectra}
\end{figure*}

The EPIC spectra were analyzed in the 0.3-10~keV energy band (corresponding to a rest-frame energy range of $\sim 2-50$~keV) with the XSPEC v.12.11.1 spectral fitting package. We performed spectral modeling adopting the Cash statistics as implemented in XSPEC with the direct background subtraction \citep[W-stat in XSPEC][]{cash1979,wachter1979}. In this energy band, the pn, MOS1 and MOS2 spectra consist of 169, 78 and 77 background subtracted counts, respectively. We modeled the X-ray spectrum with a power-law model modified by Galactic column density  of $1.14\times10^{20}~\rm cm^{-2}$ \citep[][]{HI4PI2016}, yielding a photon index of $\Gamma=1.52\pm0.11$ and a fit statistic of $W-stat=259$ for 205 degrees of freedom ($dof$). This parametrization shows a significant excess at $\sim 1-2$~keV and negative residuals elsewhere, especially at low ($<0.9$~keV) energies. Therefore, we included in the fitting model an additional  cold absorption component, parametrized by the model {\tt ztbabs} in XSPEC, accounting for the intrinsic obscuration at the quasar rest-frame. We obtained a best-fit model with $W-stat/dof=249/204$ and a better description of the data in terms of residuals (see Fig.~\ref{xrayspectra}, left panel). We found a best-fit $\Gamma=2.08_{-0.23}^{+0.27}$ with an intrinsic column density of $N_{\rm H}=1.1^{+0.6}_{-0.4}\times10^{23}~\rm cm^{-2}$, corresponding to an observed flux of 8.3$^{+2.2}_{-1.9}$ $\times$ 10$^{-15}$ erg cm$^{-2}$ s$^{-1}$ in the 2-10 keV band. 
This implies an intrinsic 2-10 keV luminosity $\log (L\rm_{X}/erg~s^{-1})=45.37_{-0.11}^{+0.13}$. Given the rest-frame high energy probed by this XMM-Newton observation, we also included an additive term accounting for Compton reflection (not modified by the QSO intrinsic absorber) due to primary coronal X-rays reflected from circumnuclear cold  material which typically peaks at 20-30~keV, by adopting the XSPEC model {\tt pexrav} \citep{MZ1995}. In this fit, we had to fix the power-law continuum slope to $\Gamma=2$ because of strong degeneracy among other free parameters. The best-fit model resulted in a negligible reflection component with reflection strength $R<0.05$ .
This is in agreement with the findings reported of other luminous QSOs, which typically show a very weak reflection component \citep[e.g.][]{reeves2000,zappacosta2018}.

The value of the X-ray continuum slope  is consistent with that expected from the $\lambda_{\rm Edd}$ $\sim$ 0.4 of SMSS J2157 according to the most recent $\Gamma$-$\lambda_{\rm Edd}$ relations \citep{liu2021,laurenti2024}. The presence of a large amount of X-ray absorption ($N_{\rm H} \approx 10^{23}~\rm cm^{-2}$) along our line of sight to the nucleus is also typically observed in the vast majority of BAL quasars at any redshift \citep[e.g.][]{gallagher2002,piconcelli2005,martocchia2017}.

Adopting the intrinsically absorbed power-law model as our fiducial best fit for the X-ray spectrum of SMSS J2157, we measured a monochromatic 2~keV luminosity of $3.16\times10^{27}\rm~erg/s/Hz$, which implies an optical-to-X-ray spectral index (a measurement of the strength of the ionizing SED) $\alpha\rm_{OX}=0.3838\log(L\rm_{2~keV}$/$L\rm_{2500\AA})=-2.03$. 
The right panel of Fig. \ref{xrayspectra} shows that this value lies below the well-established $\alpha\rm_{OX}$-$L{\rm 2500 \AA}$ relations for AGN. Specifically, we calculated a $\Delta(\alpha\rm_{OX})$ $\approx$ $-$0.2 corresponding to the difference of the $\alpha\rm_{OX}$ found for SMSS J2157 with respect to the expected value of $\alpha\rm_{OX}$ from the \citet{lusso2016} relation.
Such a $\Delta(\alpha\rm_{OX})$ is typically considered the threshold for a source to be classified as intrinsically X-ray weak \citep{luo2015,zappacosta2020}.
The X-ray bolometric correction of $K\rm^X_{Bol}$= $L\rm_{Bol}$/$L\rm_{2-10}= 2.29\times10^{48}/2.34\times10^{45} \approx 980$  is a factor of $\sim$ 2 larger than the $K\rm^X_{Bol}$ derived by assuming the $K\rm^X_{Bol}$-$L\rm_{Bol}$ relation derived by \citet{duras2020} for type 1 AGN.
Both pieces of evidence therefore lend support to consider SMSS J2157 an intrinsically X-ray weak-like QSO. This is not particularly surprising since there is growing literature that finds a sizable fraction ($\sim$30\%) of sources at the brightest end of the AGN luminosity function ($L\rm_{Bol} > 10^{47} erg~s^{-1}$) exhibiting a significantly weaker X-ray emission than other QSO at comparable   $L_{\rm Bol}$ \citep[e.g.][]{nardini2019,zappacosta2020}.

BAL QSOs tend to be weaker in X-rays than non-BAL QSOs (e.g., \citealt{Saccheo2023}), with softer SEDs, indicated by steeper $\alpha\rm_{OX}$. Another indicator of the strength of the ionizing SED is the He II$\lambda$1640 emission line, a recombination line from He III. Its intensity directly reflects the number of photons with energies above 54.4 eV, serving as a proxy for the ionizing EUV. The weakness of He II emission in SMSS J2157 spectrum (see Fig. \ref{fig:spectrum_final}) is consistent with our findings of soft X-ray continuum, which prevents overionization of the high-velocity outflow while the UV radiation accelerates the BAL outflow to high velocities (e.g. \citealt{Murray95}).
Furthermore, by recovering the CIV emission line velocity shift as derived in Sect. \ref{sec:norm}, SMSS J2157 follows the relation found by \cite{zappacosta2020} between $\alpha_{\rm ox}$ and the CIV shift, with the largest shifts exhibited by X-ray weak sources, where the launch of fast winds is favored by reduced X-ray emission.

\section{Kinematics and energetics of the EHVO outflow}\label{sec:kinematics}

\begin{figure*}
    \centering
    \includegraphics[width=9cm]{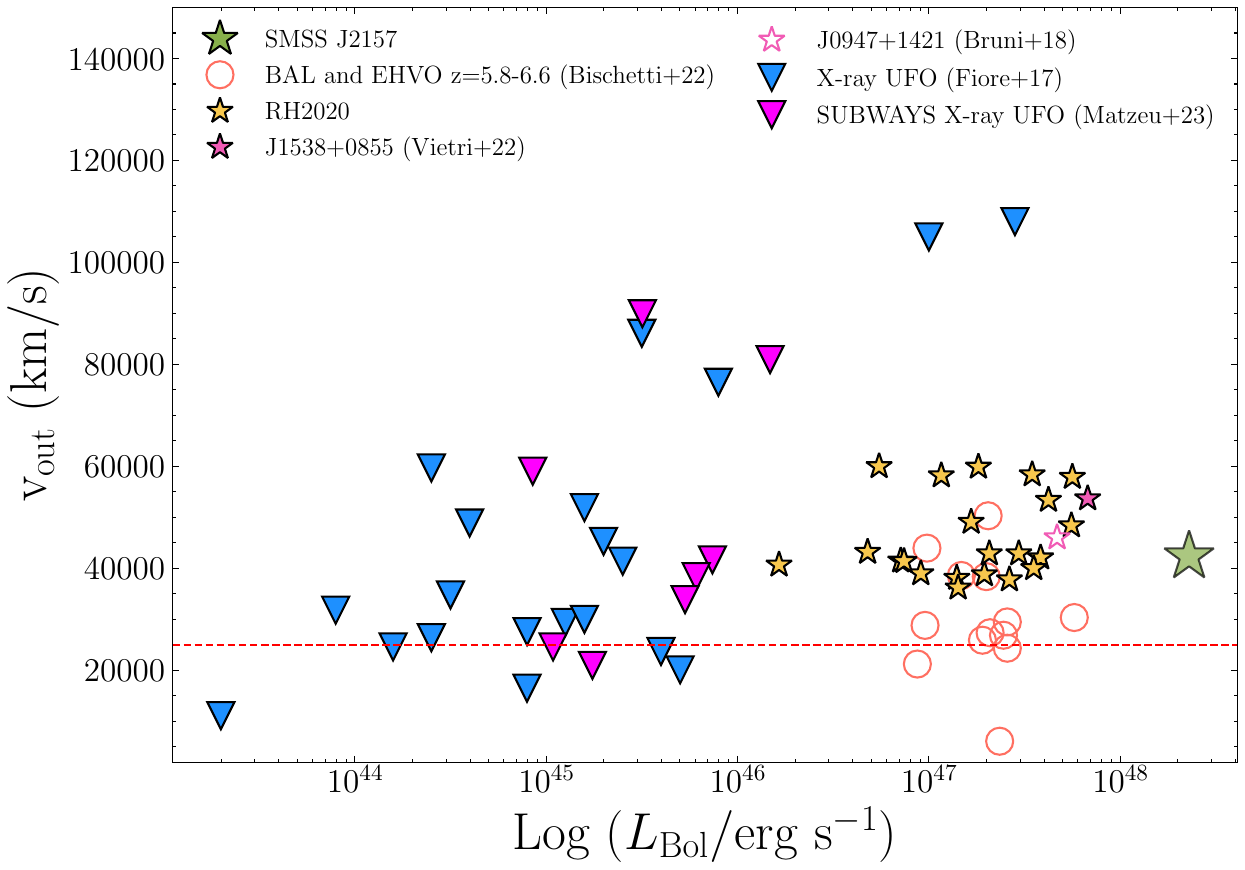}
    \includegraphics[width=9cm]{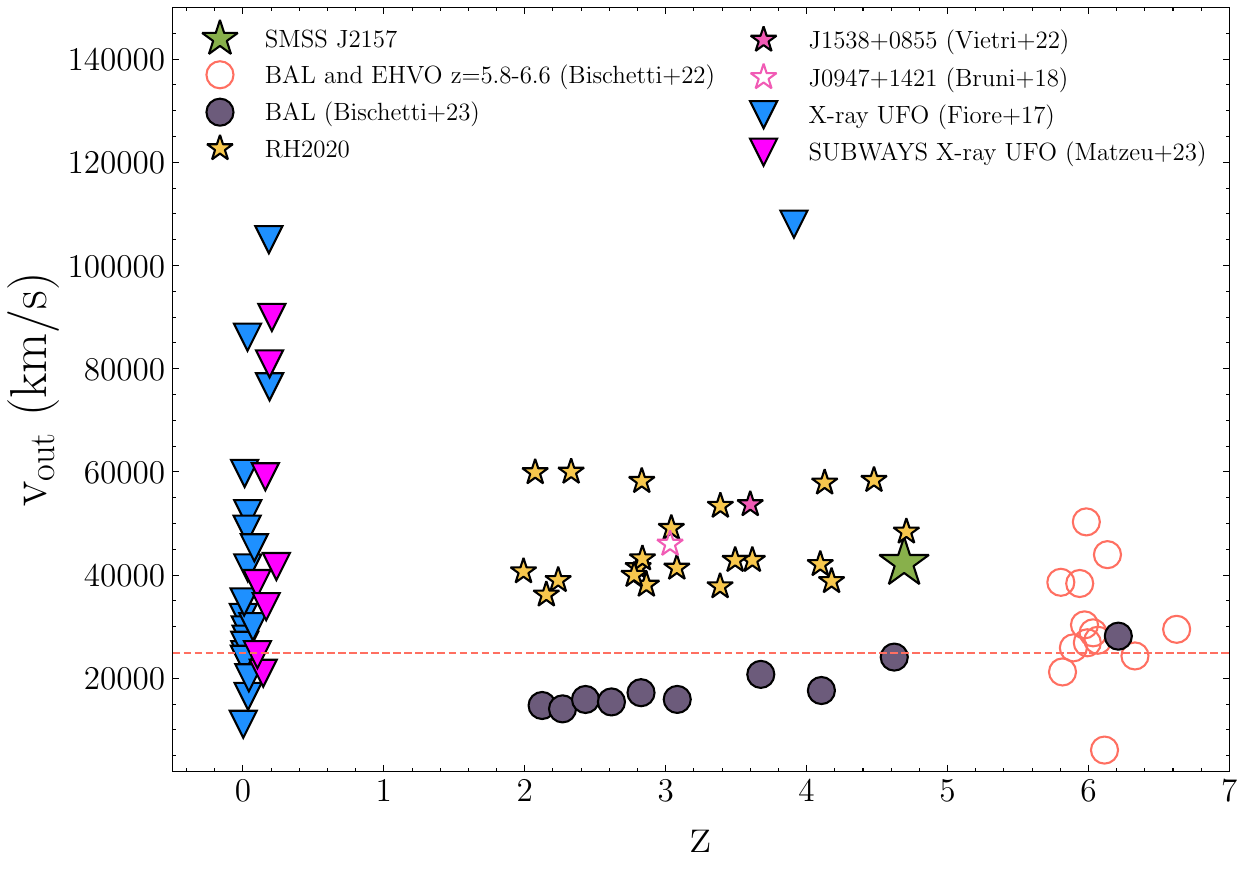}
    \caption{Velocity of different types of outflows as a function of \lbol\ (left panel) and redshift (right panel). The green star symbol denotes the EHVO outflow of SMSS J2157 with v$\rm_{out}$=v$\rm_{max}$. High-redshift BALs and EHVOs from \cite{Bischetti22} are represented as empty red circles. EHVOs from RH20 are shown as orange filled stars, while EHVOs from \cite{Vietri22} and \cite{Bruni19} are indicated by filled and empty magenta stars, respectively. The collection of X-ray UFOs from \cite{Fiore2017} and \cite{Matzeu23} (SUBWAYS sample) is reported as blue and magenta triangles, respectively. In the right panel, BAL from \cite{Bischetti23} are also shown as purple filled circles. The velocity threshold of 25,000 km/s, which distinguishes between classical low-velocity BALs and EHVOs, is shown as a red dashed line.}
    \label{fig:vmax_lbol}
\end{figure*}

Our results indicate that SMSS J2157 shows a persistent outflow with no variation over a few months to one year, with a maximum velocity of v$\rm_{max} \sim$ 0.13c. This is consistent with findings that, over such short time scales (i.e., $\sim$0.06-0.2 years), more than 90\% of BALs do not show time variation (\citealt{Capellupo13}), and the expected amplitude of variation is very small, even if they are variable (see Fig. 11 of \citealt{Gibson08}).

\citealt{Fiore2017} find a strong correlation between AGN bolometric luminosity and maximum velocity for X-ray UFOs, BALs and lower velocity X-ray absorbers. Similarly, \cite{Matzeu23} find a correlation by analyzing X-ray UFOs from the SUBWAYS sample at intermediate redshift and incorporating low- and high-redshift UFOs in their comprehensive study.
Surprisingly, no clear dependence is reported for EHVOs, as shown in Fig. \ref{fig:vmax_lbol}. EHVO data from RH2020\footnote{The gap in the velocity distribution between $\sim$25000 and $\sim$30000 km s$^{-1}$ arises from the selection strategy adopted in the search for EHVOs in RH2020, aimed at avoiding contamination from the SiIV+OIV] emission line complex.}, along with data from \cite{Bischetti22} and the WISSH QSOs J0947+1421 and J1538+0855 (\citealt{Bruni19}; \citealt{Vietri22}), are presented in the figure. Although SMSS J2157 is the most luminous source, its outflow velocity remains within the typical range observed for other EHVOs.

Notably, the parameter space for these outflows is confined to a high-luminosity regime, spanning approximately one order of magnitude, and does not extend to the highest velocities seen in X-ray UFOs (compiled from \citealt{Fiore2017} and the SUBWAYS sample from \citealt{Matzeu23}). The latter, in contrast, are observed across nearly three orders of magnitude in luminosity.

\cite{Bischetti23} observed that the velocity of classical BAL outflows (v$\le$25,000 km s$^{-1}$) generally increases with redshift, suggesting that in the high-redshift universe, BAL outflows may be more readily accelerated to EHVOs compared to later cosmic epochs, with velocities observed up to v$\sim$55,000 km s$^{-1}$ (\citealt{Bischetti22}).
Such EHVOs observed at early cosmic epochs are also found at lower redshifts, as illustrated in Fig.~\ref{fig:vmax_lbol} (right panel). This suggests that EHVOs maintain very high velocities throughout cosmic time, with values consistent with those reported at high redshift by \citet{Bischetti22}. However, a more comprehensive and systematic study of EHVOs across a wider redshift range is needed to draw more robust and reliable conclusions.

Using the physical parameters of the EHVO reported in Sect. \ref{sec:column_density}, the outflow kinetic power can be constrained with assumptions.
A key method to reduce uncertainties in these estimates is spectral variability analysis, which helps constrain the outflow distance. However, since no spectral variation is observed in the multi-epoch spectra, this method cannot be applied to estimate the distance of the EHVO from the BH. 

Therefore, we adopted the BLR radius \(R_\mathrm{BLR}\)$\sim$0.35 pc (\citealt{Lira2018}), as the location of the EHVO. This choice provides a conservative lower limit on the outflow's distance. By assuming an expanding shell at a certain velocity (v$\rm_{EHVO}$=v$\rm_{max}$) and at a radial distance $R\rm_{EHVO}$=$R\rm_{BLR}$, the kinetic energy can be expressed as:

{\small{
\begin{equation}
E_{\mathrm{K,EHVO}} = 4.8 \times 10^{53} 
\left(\frac{Q}{0.15}\right) 
\left(\frac{N_{\mathrm{H}}}{5 \times 10^{22}\ \mathrm{cm}^{-2}}\right) 
\left(\frac{R_{\mathrm{EHVO}}}{1\ \mathrm{pc}}\right)^2
\left(\frac{{\mathrm{v_{EHVO}}}}{8000\ \mathrm{km}\ \mathrm{s}^{-1}}\right)^2 
\, \mathrm{erg}
\label{eq:ekin}
\end{equation}}}

with $Q$ = 0.15 (based on the incidence of CIV BALs in SDSS QSOs, e.g. \citealt{Gibson09a}, \citealt{Hamann19}) and \NHCfmax =2.8$\times$10$^{20}$ cm$^{-2}$ and \NHCfmin = 1.1$\times$10$^{21}$ cm$^{-2}$ as derived in Sect. \ref{sec:column_density}. Dividing ${E}\rm_{K,EHVO}$ by a characteristic flow time given by $R_\mathrm{EHVO}$/v$\rm_{max}$, we find a conservative range for the kinetic power of \EkinCfmax $\sim$ 3.6 $\times$ 10$^{43}$ erg/s and \EkinCfmin $\sim$ 1.45 $\times$ 10$^{44}$ erg/s. \footnote{By adopting equations (9) and (11) from \citet{Dunn10} for $\dot{M}_{\mathrm{out}}$ and $\dot{E}_{\mathrm{K,EHVO}}$, respectively, with a mean molecular weight $\mu = 1.4$, we obtain values approximately 2.8 times higher than those derived in Sect. \ref{sec:kinematics}.}

The mass outflow rate is $\sim$ 0.06 (0.25) M$\rm{_\odot}$ yr$^{-1}$, for $C_f^{max}$ ($C_f^{min}$). Assuming a mass to radiation conversion efficiency $\eta \sim$ 0.1, the mass outflow rate corresponds to about 0.02\% (0.06\%) of the mass accretion rate ($\sim$400 M$\rm{_\odot}$ yr$^{-1}$), for $C_f^{max}$ ($C_f^{min}$).

\begin{figure}
    \centering
    \includegraphics[width=9cm]{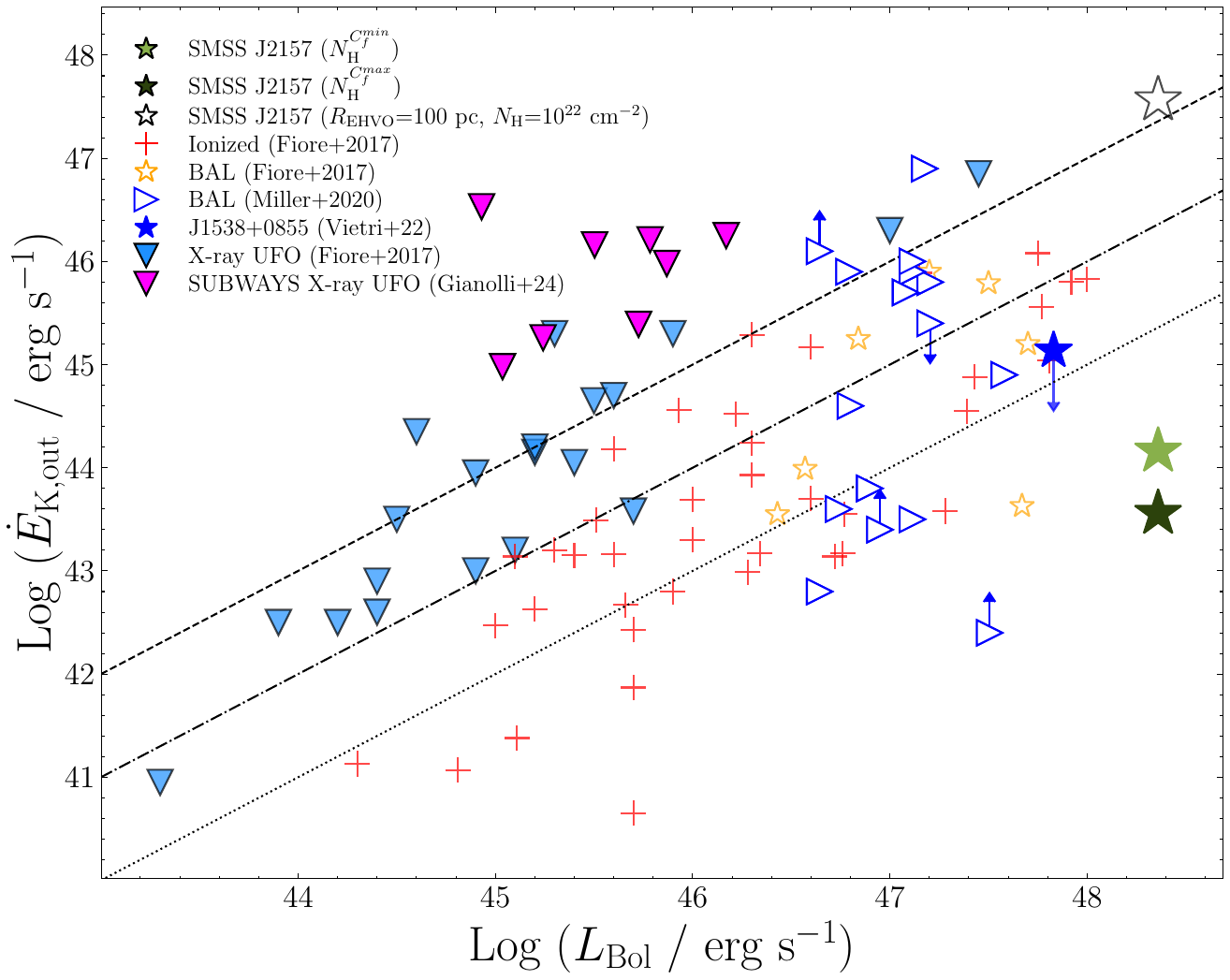}
    \caption{Distribution of $\dot{E}\rm_{K,out}$ as a function of \lbol\ for different types of outflows. The dashed, dot-dashed and dotted lines indicate the thresholds of 0.1, 1 and 10 percent, respectively. The dark and light green star symbols denote the EHVO outflow of SMSS J2157 as reported in this paper by adopting \NHCfmin and \NHCfmax, respectively, while the empty star represents $\dot{E}\rm_{K,out}$ obtained by adopting \NH\ = 10$^{22}$ cm$^{-2}$ and $R_\mathrm{EHVO}$ = 100 pc, values generally derived using VHP outflows and excited states, respectively (see Sect.~\ref{sec:kinematics} for a detailed discussion). Other symbols denote the parameters of X-ray (blue triangles), ionized (red crosses) and BAL outflow (orange stars) reported by \cite{Fiore2017}, SUBWAYS X-ray UFO from \cite{Gianolli2024}, BAL reported by \cite{Miller20} (right-pointing triangles) and EHVO from \cite{Vietri22} (blue star).}
    \label{fig:Ekin_Lbol}
\end{figure}

The kinetic power for SMSS J2157 is $\sim$ 0.002\% (0.006\%) of the bolometric luminosity, for $C_f^{max}$ ($C_f^{min}$) (see Fig.\ref{fig:Ekin_Lbol}). These values are below to what is found for X-ray UFO and BAL as found by \citealt{Fiore2017} and for X-ray UFOs from the SUBWAYS sample by \citealt{Gianolli2024}. As shown in Fig. \ref{fig:Ekin_Lbol}, about half of the X-ray absorbers and BAL winds have $\dot{E}\rm_{K}$/\lbol\ in the range 1-10\% with another half having $\dot{E}\rm_{K}$/\lbol\ $<$ 1\%. However, X-ray UFOs are usually identified in AGN with moderate luminosities ($L\rm_{Bol}\sim$10$^{43-46.5}$ erg/s), therefore under similar physical conditions (high outflow velocities and large column densities), they can more easily reach higher values of $\dot{E}\rm_{K}$/\lbol. In contrast, for very luminous quasars like SMSS J2157, the same absolute kinetic power corresponds to a much lower fraction of the bolometric output.

Moreover, various factors of uncertainty in \NH\ and $R_\mathrm{EHVO}$ can impact the outflow energetics measurement. 
\cite{Arav2013} identified two distinct ionization phases in quasar outflows, high-potential (HP) and very-high potential (VHP) outflow, observed in all outflows at $\lambda\rm_{rest}\ge$ 1050 $\AA$ and in the range 500–1050 $\lambda$ rest-frame (EUV500), respectively. The VHP exhibits an $N\rm_H$ that is 5 to 100 times larger than that of the HP \citep{Arav2020}, corresponding to $\dot{E}\rm_{K,EHVO}$ that is 1 to 2 orders of magnitude higher. Therefore, deriving the parameters of the VHP is crucial for understanding the impact of outflows on the host galaxy.
As stated in \cite{Arav2020} it is probable that the large majority of HP outflows observed at $\lambda\rm_{rest}$ $\ge$ 1050 $\AA$, as those observed in CIV line of SMSS J2157, also have an associated VHP outflow, with a derived column density which can yield measurements even larger than \NH\ = 10$^{22}$ cm$^{-2}$, more than $\sim$1 dex larger than the one derived for our target.

A robust way to derive $R_\mathrm{EHVO}$ is through the use of transitions from excited states. However these transitions are mostly observable in the EUV500 wavelength range, which is affected by Ly$\alpha$ forest at high redshift. Using this method, previous studies have found a distance of hundreds of pc (e.g. \citealt{Arav2020}), approximately three orders of magnitude larger than our conservative lower limit assuming a sub-pc BLR location. Therefore, by adopting values typically derived using VHP outflows and excited states, such as $N\rm_H$ = 10$^{22}$ cm$^{-2}$ and $R_\mathrm{EHVO}$ = 100 pc, we obtain $\dot{E}\rm_{K,EHVO}$=3.7$\times$ 10$^{47}$ erg/s ($\sim$ 16\% \lbol). 

As a result, $\dot{E}_\mathrm{K,EHVO}$ can reach values potentially capable of delivering efficient feedback to the host galaxy’s interstellar medium (based on \citet{Hopkins10}, where $\sim$0.5\% of the bolometric luminosity is typically adopted as the threshold for effective AGN feedback via outflows).

\section{Summary and conclusions}\label{sec:conclusion}
We have presented a detailed spectroscopic study of SMSS J2157, the most luminous QSO in the first 1.3 Gyr, based on multi-epoch observations with VLT/XSHOOTER, as well as combined Keck/NIRES and VLT/XSHOOTER data. In this study, we report the discovery of a persistent EHVO in SMSS J2157. Our main findings can be summarized as follows:

\begin{itemize}

 \item properties of the EHVO in absorption: the EHVO outflow displays a CIV balnicity index BI$\rm_{EHVO}^{CIV}$=2200 km s$^{-1}$, which is among the largest values discovered so far for EHVOs with velocities exceeding 35,000 km s$^{-1}$. It also shows a stable maximum velocity of v$\rm_{max} \sim$ 42,350 km s$^{-1}$ ($\sim$0.13$c$) over a monitoring period of months to a year. NV and Ly$\alpha$ EHVO components at similar velocities are also detected, although they are blended with the Ly$\alpha$ forest.

\item properties of the CIV emission line: we found an EW(CIV) $\sim$ 17 $\pm$ 2 $\AA$, placing it marginally within the weak emission line regime as defined by \citet{Chen2024}, with v$\rm_{CIV}^{50} \sim$ 4660 $\pm$ 200 km/s, indicating a strong BLR outflow. These properties are consistent with the findings of low equivalent widths and significant blueshifts in the CIV emission line observed in a sample of EHVOs analyzed by \citealt{RodriguezHidalgo22}.

\item X-ray spectral properties: SMSS J2157 exhibits a steep optical-to-X-ray spectral index (\(\alpha\rm_{OX} = -2.03\)) along with a significant X-ray bolometric correction, classifying it as a X-ray weak-like source typical of BAL QSOs. 

The source follows the $\alpha\rm_{OX}$ vs. blueshift relation for the hyperluminous quasars reported by \citealt{zappacosta2020}. The observed weakness may play a critical role in preventing overionization of the innermost disk atmosphere, hence allowing an efficient launch of the fastest nuclear UV outflows such as the EHVO presented in this work.

\item energetics of the EHVO: our estimates place a conservative range of values of the kinetic power of the outflow at \EkinCfmax $\sim$ 3.6 $\times$ 10$^{43}$ erg/s and \EkinCfmin $\sim$ 1.45 $\times$ 10$^{44}$ erg/s, for $C_f^{max}$ and $C_f^{min}$, respectively, which, given the extremely high bolometric luminosity of SMSS J2157, corresponds to approximately 0.002\% (0.006\%) of the quasar’s radiative output, for $C_f^{max}$ ($C_f^{min}$), significantly below the 0.5-5\% of bolometric luminosity threshold typically considered necessary for efficient AGN feedback mechanisms. However, this value reflects conservative assumptions on the outflow location and uncertainties in the column density (see Sect. \ref{sec:kinematics}).
    
\end{itemize}
Our findings suggest that EHVOs may be a common feature of luminous QSOs across all redshifts and, in turn, could provide an efficient feedback mechanism for the co-evolution of galaxies hosting highly-accreting, massive SMBHs.
Future monitoring campaigns are needed to confirm the persistence of the outflow over timescales of years. Additionally, variability, if detected, can provide stringent constraints on the outflow's location, ultimately reducing uncertainties in the kinetic power estimates.


\begin{acknowledgements}
We thank the anonymous referee for the useful comments that improved the paper. We warmly thank Vittoria Gianolli for kindly providing us with data from the SUBWAYS sample. Based on observations collected at the European Southern Observatory under ESO programme 0103.B-0949(A). G. V. acknowledges financial support from the Bando Ricerca Fondamentale INAF 2022 Mini-grant "Searching for UV ultra-fast outflow in AGN by exploiting widearea public spectroscopic surveys" and INAF 2023 Guest Observer Grant "Assessing the role of ultra-fast outflows in hyper-luminous
quasars at Cosmic Noon". P.R.H. and L.F. acknowledge support from the National Science Foundation AAG Award AST-2107960, the Sloan Digital Sky Survey's Faculty And Student Team program, funded by the Alfred P. Sloan Foundation, and the Mary Gates research scholarship program. EP acknowledges funding from the European Union - Next Generation EU, PRIN/MUR 2022 2022K9N5B4. T.M. acknowledges support from JSPS KAKENHI Grant Number 25K01038. ALR acknowledges support from a UKRI Future Leaders Fellowship (grant code: MR/T020989/1). LZ acknowledges financial support from the Bando Ricerca Fondamentale INAF 2022 Large Grant “Toward an holistic view of the Titans: multi-band observations of z > 6 QSOs powered by greedy supermassive black holes” and from the European Union – Next Generation EU, PRIN/MUR 2022 2022TKPB2P – BIG-z. For the purpose of open access, the author has applied a Creative Commons Attribution (CC BY) licence to any Author Accepted Manuscript version arising from this submission. 

\end{acknowledgements}


\bibliographystyle{aa} 
\bibliography{bibliography_new} 

\begin{appendix}

\section{CIV blended-NALs}\label{app:app}

\begin{figure}[H]
    \includegraphics[width=8cm]{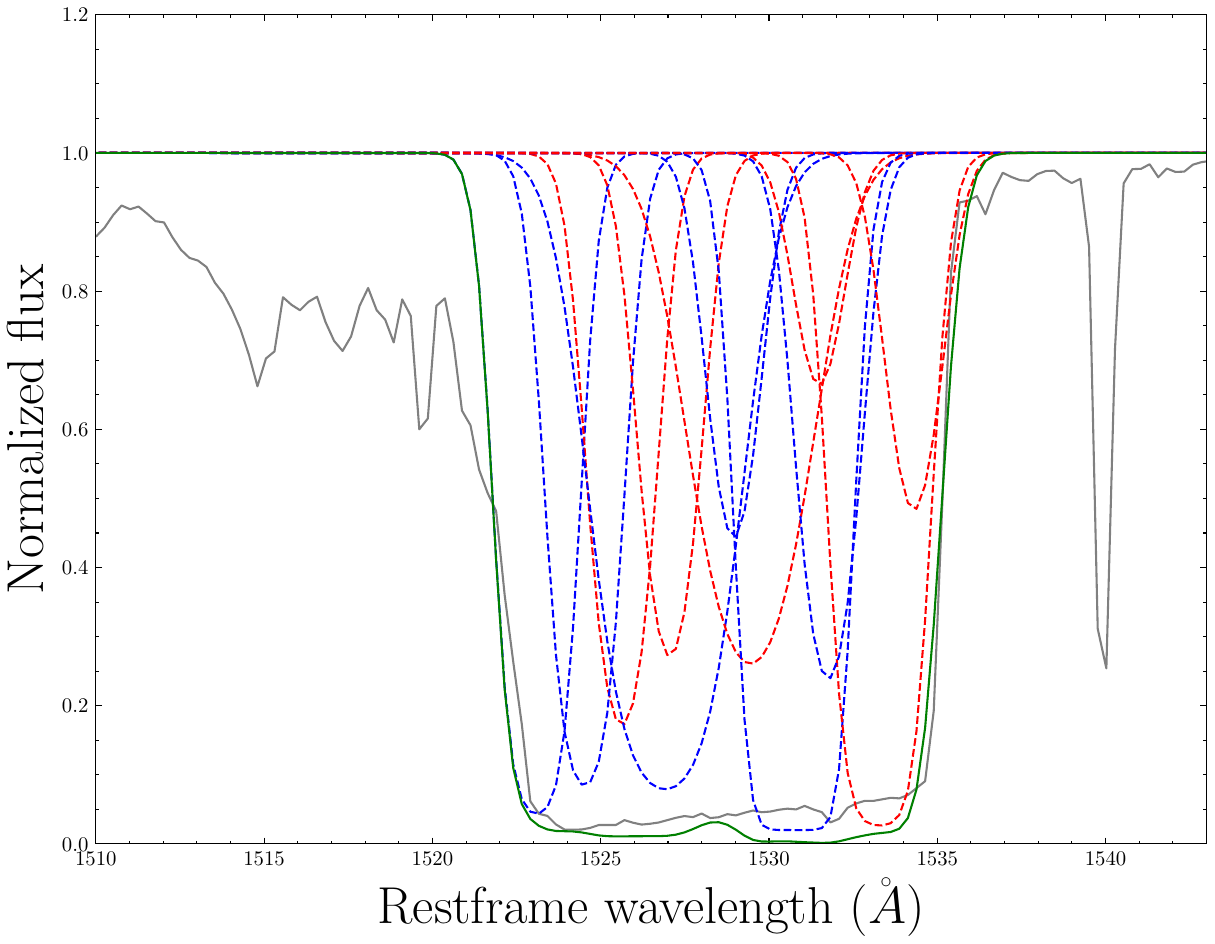}
    \caption{Normalized coadded spectrum of SMSS J2157 in the wavelength range of the CIV blended-NALs shown together with its best-fit model (green curve) and six pairs of Gaussian components (blue and red) modeling the doublets (see Sect. \ref{sec:bal} for more details). 
    }
 
    \label{fig:CIV_blended}
\end{figure}

\end{appendix}

\end{document}